\documentstyle[12pt,epsfig]{article}

\advance\textheight by \topskip
\newcommand{\be}{\begin{equation}}
\newcommand{\ee}{\end{equation}}
\newcommand{\br}{\begin{eqnarray}}
\newcommand{\er}{\end{eqnarray}}
\newcommand{\ba}{\begin{array}}
\newcommand{\ea}{\end{array}}
\newcommand{\bi}{\begin{itemize}}
\newcommand{\ei}{\end{itemize}}
\newcommand{\bn}{\begin{enumerate}}
\newcommand{\en}{\end{enumerate}}
\newcommand{\bc}{\begin{center}}
\newcommand{\ec}{\end{center}}

\newcommand{\Dir}{\kern -6.4pt\Big{/}}
\newcommand{\Dirin}{\kern -10.4pt\Big{/}\kern 4.4pt}
\newcommand{\DDir}{\kern -8.0pt\Big{/}}
\newcommand{\DGir}{\kern -6.0pt\Big{/}}

\def\slashchar#1{\setbox0=\hbox{$#1$}           
     \dimen0=\wd0                                 
     \setbox1=\hbox{/} \dimen1=\wd1               
     \ifdim\dimen0>\dimen1                        
        \rlap{\hbox to \dimen0{\hfil/\hfil}}      
        #1                                        
     \else                                        
        \rlap{\hbox to \dimen1{\hfil$#1$\hfil}}   
        /                                         
     \fi}                                         %

\def\be{\begin{equation}}
\def\ee{\end{equation}}
\def\bea{\begin{eqnarray}}
\def\eea{\end{eqnarray}}
\def\lsim{\:\raisebox{-0.5ex}{$\stackrel{\textstyle<}{\sim}$}\:}
\def\gsim{\:\raisebox{-0.5ex}{$\stackrel{\textstyle>}{\sim}$}\:}

\def\pl #1 #2 #3 {{Phys.~Lett.} {#1} (#2) #3}
\def\np #1 #2 #3 {{Nucl.~Phys.} {#1} (#2) #3}
\def\zp #1 #2 #3 {{Z.~Phys.} {#1} (#2) #3}
\def\jp #1 #2 #3 {{J.~Phys.} {#1} (#2) #3}
\def\pr #1 #2 #3 {{Phys.~Rev.} {#1} (#2) #3}
\def\prep #1 #2 #3 {{Phys.~Rep.} {#1} (#2) #3}
\def\prl #1 #2 #3 {{Phys.~Rev.~Lett.} {#1} (#2) #3}
\def\mpl #1 #2 #3 {{Mod.~Phys.~Lett.} {#1} (#2) #3}
\def\rmp #1 #2 #3 {{Rev. Mod. Phys.} {#1} (#2) #3}
\def\cpc #1 #2 #3 {{Comp. Phys. Commun.} {#1} (#2) #3}
\def\sjnp #1 #2 #3 {{Sov. J. Nucl. Phys.} {#1} (#2) #3}

\def\slash{/\kern -5pt}

\def\ims #1 {\ensuremath{M^2_{[#1]}}}

\def\s22w{s_{2W}^2}

\begin{document}
\tolerance=100000
\thispagestyle{empty}
\setcounter{page}{0}
 \begin{flushright}
{\large CERN-TH/2001-036}\\
{\large April 2002}\\
\end{flushright}

\vspace*{\fill}

\begin{center}
{\Large \bf Pair production of charged Higgs scalars\\[0.5cm]
from electroweak gauge boson fusion}\\[2.cm]

{\Large S. Moretti}\\[5mm] 
{Theory Division, CERN, CH-1211 Geneva 23,
Switzerland} 

\end{center}

\vspace*{\fill}

\begin{abstract}{\small\noindent
We compute the contribution to charged Higgs boson pair production
at the Large Hadron Collider (LHC) due to the scattering of
two electroweak (EW) gauge bosons, these being in turn
generated via bremsstrahlung off incoming quarks:
$q q\to q q V^*V^*\to q q H^+H^-$ ($V=\gamma,Z,W^\pm$).
We verify that the production cross section of this mode 
is $\tan\beta$ independent and show that it is
smaller than that of $H^+H^-$  production via 
$q q$-initiated processes
but generally larger than that of the loop-induced channel $gg\to H^+H^-$.
Pair production of charged Higgs bosons is crucial in order to test EW
symmetry breaking scenarios beyond the Standard Model (SM). 
We show that the detection of these kind of processes at the standard 
LHC is however problematic, because of their poor production rates and 
the large backgrounds.}
\end{abstract}

\vspace*{\fill}
\newpage
\setcounter{page}{1}
\section{Introduction}
\noindent
A charged scalar state does not belong to the particle spectrum
of the SM. Therefore, to detect a signal of it
would definitely confirm the existence of New Physics.
A framework that can naturally accommodate such a particle
is the Minimal Supersymmetric Standard Model (MSSM). This is a realisation
of a general two Higgs Doublet Model (2HDM) within the theoretical
framework provided by Supersymmetry (SUSY). 
While the SM incorporates only one `neutral' Higgs boson, $\phi$, the
MSSM  predicts a pair of `charged' Higgs bosons $H^\pm$
along with three `neutral' ones, the CP-even  $H$ and $h$ and the
CP-odd  $A$ \cite{guide}. 

It is then not surprising the considerable interest, that has revived
lately \cite{theory,experiment}, in accessing the Higgs sector of the 
MSSM at the 
future CERN collider through the detection of charged Higgs boson states.
In fact,  one may well conjecture that, even in presence of a clear signal of a
neutral Higgs particle,
it could be difficult to distinguish between the SM and the lightest MSSM
Higgs boson. For example, in the so-called `decoupling regime' of the MSSM, 
 one has that the $h$ couplings to ordinary
matter become similar to those of the SM $\phi$ and, besides,
the MSSM Higgs masses are such that $ M_h\ll M_H\approx 
M_A\approx M_{H^\pm}$. In fact, such a decoupling scenario occurs
for $M_{H^\pm}\gsim 150-200$ GeV, for any value of $\tan\beta$\footnote{Here,
$\tan\beta$ denotes the ratio of the vacuum expectation values of the 
two Higgs doublets of the MSSM and can conveniently be used to parametrise 
at tree level the entire Higgs sector, alongside one
of the masses, e.g., $M_{H^\pm}$ itself.}. Under these circumstances,
it would probably be equally challenging to detect a second neutral 
Higgs signal, as it would be to select a charged Higgs boson
signature. This scenario, though
not to be expected necessarily, may be viewed as not at all
unreasonable,  especially taking into account the latest LEP2 results on 
the possible existence of a neutral Higgs state with mass 
of about 110--115 GeV  \cite{three} ($M_{h}$ in the MSSM), 
 which, using the MSSM Higgs mass relations (now, known 
at two-loops \cite{twoloop}), implies an indirect lower 
bound on $M_{H^\pm}$ already at 140 GeV in the low $\tan\beta$ region,
say, around $3$ or so\footnote{The direct experimental limit obtained 
at LEP2 on $M_{H^\pm}$, based
on searches for $e^+e^-\to H^+H^-$ events, is at present much 
smaller: just below $M_{W^\pm}$ \cite{three}.}.  

At the LHC, light charged Higgs scalars
(i.e., with $M_{H^\pm}<m_t$) can be produced either in top decays,
$t\to bH^+$ (with the top quarks being mainly produced via $gg\to t\bar t$),
or in pair from quark-antiquark annihilations,
$q\bar q\to H^+H^-$ \cite{qqHH}\footnote{At the forthcoming 
Run II of the upgraded Fermilab 
Tevatron, the first of these channels will allow experimenters to scan the
$[M_{H^\pm},\tan\beta]$ plane for large and small values of
$\tan\!\beta$, say, below 2 and above $m_t/m_b$,
roughly up to the kinematical limit of the
$t \rightarrow b H^+$ decay, $M_{H^\pm}\approx 
m_t - m_b$~\cite{Tevatron}, whereas
the second channel will be useful
in the intermediate $\tan\!\beta$ region \cite{Kosuke}, provided
charged Higgs bosons are light enough, as  simple phase space suppression 
severely handicaps pair production at $\sqrt s=2$ TeV.}. 
Heavy charged Higgs bosons (i.e., with $M_{H^\pm}>m_t$ -- those
beyond the reach of the Tevatron) are mainly generated
via the reaction $g\bar b \rightarrow \bar tH^+$~\cite{bg}.
(In fact, the two processes, $gg\to t\bar t$, with $t\to bH^+$,
and $g\bar b \rightarrow \bar tH^+$, can be connected \cite{Borz,mono}
by looking at the generic subprocess  $gg \rightarrow \bar t{b}H^+$~\cite{gg}.)
Alternative production channels at the LHC are, for heavy $M_{H^\pm}$:
the $bq \rightarrow bH^\pm q'$
mode of Ref.~\cite{bq}; again, charged Higgs boson pair production, but
now supplemented by the loop-induced subprocesses  
$gg \rightarrow H^+H^-$ \cite{gghh1}--\cite{YWLMH}; 
associated production $gg,b\bar{b} \rightarrow W^{\pm}H^{\mp}$ 
\cite{WH} and $q\bar q'\to \Phi H^\pm$, with $\Phi=h,H$ and $A$
\cite{qqHH}.

If one makes the assumption that the typical mass scale of sparticle states 
is much higher than $M_{H^\pm}$ 
(e.g., $M_{\mathrm{SUSY}}=1$ TeV -- as we do throughout the paper), 
thus preventing the
decay of charged Higgs bosons via SUSY channels, the decay signature
of $H^\pm$ bosons is fairly model independent
and dominated by four decay channels at most \cite{BRs}:
$H^+\to t\bar b$, $H^+\to \tau^+\nu_\tau$, $H^+\to c\bar s$ 
and $H^+\to W^+h$. By exploiting the above production channels 
in conjunction with these decay modes, it has been shown 
that $H^{\pm}$ scalars with masses up to $400\,
\hbox{GeV}$ can be discovered at the LHC if 
$\,\tan\!\beta \, \lsim \, 3$ (which is in the
neighborhood of the indirect limit from LEP2) {or} 
$\tan\!\beta \, \gsim \, 10-15$ (with the minimum occurring
when $M_{H^{\pm}}$ is close to 
$m_t$). Alternatively, if one allows for the contribution of
SUSY decay modes, according to Ref.~\cite{HSUSY}, the
surviving region $3\lsim \tan\beta \lsim10$ 
(with $M_{H^\pm}{\lsim}400\,\hbox{GeV}$)
should adequately be covered by resorting to the decays 
$H^{\pm} \rightarrow {\tilde\chi}_1^{\pm}{\tilde\chi}_1^0$ and
$H^{\pm} \rightarrow {\tilde\chi}_1^{\pm}{\tilde\chi}_2^0,
{\tilde\chi}_1^{\pm}{\tilde\chi}_3^0$, i.e., into
combinations of chargino-neutralino pairs\footnote{According
to the typical $H^\pm$ decay rates found in \cite{HSUSY},
one should not expect the scope of the SM channels 
to be spoiled by the presence of the new SUSY modes.}.

Once charged Higgs bosons will have been detected through
the leading production channels, and their mass measured, the emphasis
will turn to studying their properties. In fact, with the high luminosity
of the LHC, also the various subleading production processes would be
established experimentally. Among these, it is the $gg\to H^+H^-$
channel, originally discussed in Ref.~\cite{gghh1},
that has gathered considerable attention in the recent years
\cite{gghh,YWLMH}. The reason is twofold. Firstly, it in principle 
allows one to measure directly the strength of the trilinear 
vertices between charged and neutral
CP-even Higgs bosons, $hH^+H^-$ and $HH^+H^-$, thus
shading light on the structure of the 
Higgs sector of the MSSM\footnote{As a result of
CP-invariance, there exists no $AH^+H^-$ coupling at tree level.}. The
determination of these couplings is in fact
a necessary step in reconstructing the
self-interaction terms in the full Higgs potential. In contrast, in all other
processes mentioned, only gauge ($\gamma H^+H^-$, $Z H^+H^-$, 
$h W^+H^-$, $H W^+H^-$ and $A W^+H^-$) and Yukawa ($t\bar b H^-$) 
couplings can be accessed. Secondly, effects of SUSY could be manifest
in the $gg\to H^+H^-$ channel, because of virtual squark loops
entering the production stage,
even when charged Higgs bosons are too light to decay directly into
sparticles. Unfortunately, from these points of views,
the $gg\to H^+H^-$ subprocess is biased by the presence
of the $q\bar q\to H^+H^-$ channel, whose production rates are much
larger at the LHC,
and a complete phenomenological analysis on how to 
disentangle the two is still lacking.

\section{Pair production of charged Higgs bosons}
\noindent
The purpose of this paper is to show that, at the LHC, pairs of charged Higgs
bosons can also be produced via the fusion of gauge vector bosons,
$\gamma$, $Z$ and $W^\pm$, at rates comparable to those induced
by parton fusion in $q\bar q$ and $gg$ scatterings. Here, the gauge bosons
are emitted by incoming quarks and -- 
to a somewhat lesser extent (because of the proton-proton scattering) -- 
by antiquarks too, and the all process can
be sketched as
follows:
\be\label{signal}
q q \to q q V^*V^* \to q q H^+H^-\qquad(V=\gamma,Z,W^\pm),
\ee
where $q$ refers to both quarks and antiquarks (of any possible flavour) in the
appropriate combinations\footnote{For simplicity, we have taken the
Cabibbo-Kobayashi-Maskawa mixing matrix to be diagonal.}.
The Feynman diagrams involved can be found in Fig.~\ref{fig:graphs}.
However, notice that not all of 
these appear for each quark flavour combination.
(For example, for $u \bar u\to u\bar u H^+H^-$ no $W^\pm$ mediated diagrams
enter whereas for $u \bar u\to d\bar d H^+H^-$ the latter are needed.)
Besides, we have not shown the graphs that differ from those depicted
in Fig.~\ref{fig:graphs} only in the exchange of a fermion leg, as it 
happens when
identical flavours appear in the initial and final states
and we have neglected those in which Higgs bosons are radiated
by the quark lines, because of the small Yukawa couplings of the
leading valence quarks. The matrix element (ME) of process (\ref{signal})
has been computed by means of helicity amplitude techniques.
Furthermore,
it has been checked for gauge and BRS invariance and integrated numerically
over a four-body phase space.

The total cross section for process (\ref{signal}) at the
LHC ($\sqrt s=14$ TeV), before any acceptance cuts, can be found in 
Fig.~\ref{fig:ggHH} (continuous line), compared to
the yield of the other two production processes of charged Higgs boson
pairs, i.e., $q\bar q \to H^+H^-$ and $gg \to H^+H^-$,
for three reference values of $\tan\beta$ and 
with $M_{H^\pm}$ in the range $130$ to 400 GeV\footnote{For
reference, hereafter, we use the MRS(LO05A) \cite{MRS}
Parton Distribution Functions
(PDFs), with factorisation/renormalisation scale 
$Q=\mu={\sqrt{\hat{s}}}$, i.e., the centre-of-mass
(CM) energy at parton level. Other choices, such as $Q=\mu=p_T^{H^+H^-}$,
yield differences of the order of a few percent.}. 
Notice that in order to obtain
a finite answer for process (\ref{signal}) (in the case of photon exchange), 
we have adopted a non-zero value for the mass of all quark flavours.
We have chosen $m_u=m_d=0.32$ GeV, $m_s=0.50$ GeV, $m_c=1.55$ GeV and
$m_b=4.25$ GeV. As for the mass and couplings of the neutral Higgs bosons,
we have used the Renormalisation Group 
(RG) improved one-loop relations of Ref.~\cite{twoloop1}. (Notice that
mixing effects are here irrelevant, i.e., non-zero values of the soft SUSY 
parameters  
$\mu$, $A_b$ and $A_t$ entering the definition of Higgs masses and couplings
mainly affect the neutral Higgs sector while having 
negligible impact on the phenomenology of pair production of `heavy' charged
Higgs bosons at the LHC). 

Contrary to the $q\bar q\to H^+H^-$ and $gg\to H^+H^-$ channels,
process (\ref{signal}) shows no visible $\tan\beta$ dependence,
because of the tiny contribution of the $H,h$ and $A$ mediated graphs 
(numbers 15,16 and 18 in Fig.~\ref{fig:graphs}).
In the first process, such a dependence is induced mainly
by $b\bar b\to H^+H^-$
scatterings, whereas in the second one, it enters via both the triangle
and the box quark-graphs: see Fig.~1 of Ref.~\cite{YWLMH}\footnote{Here,
for simplicity, we are not including the effects of squark loops
in $gg\to H^+H^-$: again, see Fig.~1 of \cite{YWLMH}. These
can significantly enhance the corresponding cross section, e.g.,
by up to 50\%, depending upon $\tan\beta$,
in a Minimal Supergravity (MSUGRA) inspired scenario.}.
The overall rate of process (\ref{signal}) is generally larger than
that of $gg\to H^+H^-$ (except for very large $\tan\beta$ values) 
but smaller than that of  $q\bar q\to H^+H^-$,
though asymptotically (i.e., for very large values of $M_{H^\pm}$),
$q q\to q qH^+H^-$ approaches $q\bar q\to H^+H^-$.
For an annual integrated luminosity of 100 inverse femtobarns,
something like 1500 to 150 events of the type (\ref{signal})
could be produced per experiment at the LHC, for charged Higgs boson masses 
ranging from 140 to 400 GeV.

A word of caution should be spent here though, concerning the simulation
of the $b\bar b$ component of the $q\bar q\to H^+H^-$ process. In fact,
the use of a `phenomenological' $b$-quark parton density, as
available in most PDF sets currently on the market, requires crude 
approximations of the partonic kinematics, which result in a
mis-estimation of the production cross section. (The problem is well
known already from the study of the leading production processes
of charged Higgs bosons at the LHC, namely,
$g\bar b \rightarrow \bar tH^+$ and $gg\to b\bar t H^+$:
see, e.g., \cite{Borz,Scott}.) In practise, the $b$-(anti)quark in
 the initial state 
comes from a gluon in the proton beam splitting into a collinear
$b\bar b$-pair, resulting in large factors of $\sim\alpha_s \log(Q/m_b)$,
where $Q$ is the factorisation scale. These terms are 
then re-summed to all orders, $\sum_{n}\alpha_s^{n} \log
^{n}(Q/m_b)$, in evaluating the phenomenological $b$-quark PDF.
In contrast, in using a gluon density for
\begin{equation}\label{ggbbHH}
gg\to b\bar b H^+H^-
\end{equation}
 (see Fig.~\ref{fig:ggbbHH}
for the associated Feynman graphs), one basically only includes the first terms of
the corresponding two series, when the $b$ and $\bar b$ in the final state are
produced collinearly to the incoming gluon directions. In turns out that,
for $Q\gg m_b$, as it is the case here (owing to the presence
of two large masses in the final state, so that $Q\gsim 2M_{H^\pm}$),
the re-summed terms are large and over-compensate the contribution of 
the large transverse momentum, or $p_T$,
 region available in the gluon-induced case.
Fig.~\ref{fig:bbHH} illustrates
this. There, we have plotted the $b\bar b \to H^+H^-$ cross
section against that for $gg\to b\bar b H^+H^-$, for the usual choice 
PDFs, scales, $\tan\beta$ and $M_{H^\pm}$ values
as in the previous plots. For a start, one should appreciate,
by comparing Fig.~\ref{fig:ggHH} to \ref{fig:bbHH}, that $b\bar b$
rates are subleading with respect to the $q\bar q$ ones, which also
include annihilation via $u,d,s$ and $c$-(anti)quarks, even at very large
values of $\tan\beta$, where one might expect the combined effects
of the Yukawa couplings entering the $t\bar b H^-$ and $b\bar b H/h$
vertices to be largest. Moreover, in the heavy mass range,
differences between the two cross sections
can be even larger than an order of magnitude, well
in line with the findings of Refs.~\cite{Borz} and \cite{3b}, if
one considers that two $g\to b\bar b$ splittings are involved here.
(The sudden rise at $M_{H^\pm}\lsim m_t-m_b$ seen for the
gluon-induced process is due to intermediate top-antitop production,
taking place via graphs 1,5,9,10,11,15,16,20,24,25,26,30,31,35 and 39
of Fig.~\ref{fig:ggbbHH}.) Despite a well defined procedure exists to
combine the $b\bar b$- and $gg$-initiated processes, through the subtraction
of the common logarithm terms (see \cite{Borz}), we refrain from
doing so here, as Fig.~\ref{fig:bbHH} is presented with the sole
intention of making
clear that current predictions of $b\bar b$ contributions to the
production of charged Higgs bosons using
$b$-quark densities (see, e.g., some of the results in Ref.~\cite{gghh1,WH})
may be too optimistic.

Fig.~\ref{fig:comparison} presents the LHC 
cross section of the single $H^\pm$-production mechanisms 
discussed above, for $M_{H^\pm}\gsim m_t$ and 
our usual three choices of $\tan\beta$. (Here, we have plotted 
$gg\to \bar t b H^+$, accounting for the
aforementioned top-antitop production and decay
as well as the $bg$ fusion channel, with
the mentioned subtraction term included.)
Clearly, by comparing Fig.~\ref{fig:ggHH} to Fig.~\ref{fig:comparison} 
(notice the different normalisation), one realises that
process (\ref{signal}) is never dominant, although,
at $\tan\beta=7$, it is just above one order of magnitude
smaller than the dominant $gg\to \bar t b H^+$ mode.
The reason of the drop in production rates of the latter
process, similarly to what happens for $bq\to b H^\pm q'$
and $b\bar b,gg\to W^\pm H^\mp$, see Fig.~\ref{fig:comparison},
is due to a coupling of the form
\begin{equation}\label{coupling}
\sim\frac{g}{2{\sqrt{2}}M_{W^\pm}}H^+ \left( 
m_t \cot\!\beta \bar{t} b_L + m_b \tan\!\beta \bar{t} b_R
\right),
\end{equation}
whose square -- entering the corresponding 
production cross sections -- has a minimum at 
$\tan\!\beta \simeq 7$ (this is indeed the reason
of the similar  trend seen in the previous figures for 
$b\bar b\to H^+H^-$,
$gg\to b\bar b H^+H^-$
and $gg\to H^+H^-$).

The independence of $\tan\beta$
is an attractive feature that could in principle render 
process (\ref{signal}) an interesting discovery channel of
charged Higgs bosons, complementary to all other 
modes proportional to the square of the expression 
 in eq.~(\ref{coupling})\footnote{The $\tan\beta$ dependence 
of the  $q\bar q'\to \Phi H^\pm$ processes, with $\Phi=h,H$ and $A$,
is less straightforward than in eq.~(\ref{coupling}), as
all these channels also proceed via $\Phi q\bar q$,
$\Phi W^\pm H^\mp$ and $\Phi H^\pm H^\mp$ vertices, 
which are more complicated functions of $\tan\beta$, involving
in particular the Higgs `mixing angle' $\alpha$.}.
In fact, a simple measurement of the
total cross section $\sigma(q q\to q q H^+H^-\to q q X)$,
above the SM rates, would suffice to estimate an
$M_{H^\pm}$ value, that could then be employed in background
suppression in some suitable Higgs decay channel. 
In order to attempt a Higgs mass reconstruction, we
proceed as follows. First, we exploit the 
presence of two forward/backward jets in the final state
of process (\ref{signal}), that can be used
for tagging purposes and QCD background suppression, pretty much in the
same spirit as in Ref.~\cite{dieter1}
(see also \cite{dieter2}). There, it was shown how, in the SM Higgs process
\be\label{SMsignal}
q q \to q q V^*V^* \to q q \phi\to q q W^+W^-,
\ee
proceeding via $W^+W^-$ and $ZZ$ fusion ($V=Z,W^\pm$), 
the selection of the two (rather forward/backward) 
quark-jets in the final state,
within a detector acceptance region defined by $p_T^j>20$
GeV and $|\eta^j|<5$, can aid to strongly reduce the 
overwhelming (but rather central) QCD background in $t\bar tjj$
and $W^+W^-jj$ events, where $j$ represents a jet,
thus rendering process (\ref{SMsignal}) a viable
mechanism to detect SM Higgs signals via leptonic $W^+W^-$ 
decays\footnote{In fact, the decay $W^+W^-\to \ell^+\ell^-
\nu_\ell\bar\nu_\ell$ represents the best way to extract the SM 
Higgs signal from the mentioned backgrounds \cite{herbi,tw}
over the mass region between 130 and 180 GeV or so.}.
The cross section of process (\ref{signal}), after
the above transverse momentum and pseudorapidity cuts
are enforced, is shown in Fig.~\ref{fig:all} (dashed line). 
The loss of signal events, with respect to the total rate
(continuous line), is rather contained (around 35\%, typically), owning
to the fact that, for  $M_{H^\pm}\gsim 130-140$ GeV,
the bulk of the production rates is due to 
$W^+W^-$ fusion (followed by $ZZ$) rather than to $\gamma\gamma$ 
(or even $\gamma Z$), 
which could be relevant only at very small values of $M_{H^\pm}$, 
as one can deduce from Fig.~\ref{fig:contributions}. 
(Notice the peak at $M_{VV}\approx M_{Z}$, due to the `resonant'
sub-scattering $W^{+*}W^{-*}\to Z\to H^+H^-$, since we have
used  $M_{H^\pm}=10$ GeV as an illustration, value for which the $\gamma\gamma$
contribution is similar in size to the one induced by all the other graphs.)
This tendency can already be
appreciated by a simple integration over the four-body final state
of the two (squared) vector boson propagators in Fig.~\ref{fig:graphs}. 
However, notice that at very low $M_{H^\pm}$ values,
despite being numerically stable, the rates for the photon-exchange 
contributions obtained by using the quark masses as regulators of the 
otherwise divergent collinear configurations 
(between initial- and final-state quarks, 
when the photon is nearly on-shell) are not to be trusted,
as these singularities have to be subtracted and absorbed into an
additional contribution of the non-perturbative (recall that
the masses of $u,d$ and $s-$ quarks are below 1 GeV)
photon spectrum inside the proton. Nonetheless we believe
that they serve well the purpose of justifying a posteriori 
our initial calculation of process (\ref{signal}) without any 
transverse momentum 
cuts in the forward/backward jets, for the mass region of interest
here, $M_{H^\pm}\gsim m_t$, where the $\gamma\gamma$ contribution
turns out to be negligible. In fact, the final rates presented below
for the signal, after our selection procedure, have been obtained
in presence of $p_T$ cuts on the various jets, which remove
entirely the mentioned singularities.

A tentative selection procedure of the mass resonance
in the signal could be the one sketched
below. Notice that we carry out our
analysis at parton level only, thus neglecting
parton shower and hadronisation effects, although we account for 
typical detector resolutions, as
the transverse momenta of all visible particles in the final state have
been smeared according to a Gaussian distribution,
with $(\sigma(p_T)/p_T)^2 = (0.6/\sqrt{p_T})^2 + (0.04)^2$ for all
jets and
$(\sigma(p_T)/p_T)^2 = (0.12/\sqrt{p_T})^2 + (0.01)^2$ for the leptons.
The missing transverse momentum has been evaluated from
the vector sum of the jet and lepton transverse momenta after 
resolution smearing.  

\begin{enumerate}
\item We ask for the decays $H^+\to t\bar b\to b\bar b W^+$
and $H^-\to\tau^-\bar\nu_\tau$, and charge conjugate cases,
with the $W^+$ decaying hadronically to a pair of light jets.
At the same time, one requires to tag the $\tau^-$  
via its leptonic or hadronic decay channels\footnote{We assume
that the latter can efficiently be distinguished from
the shower of an (anti)quark or gluon.}. Hence, the
final signature is 
`$6~{\mathrm{jets}}~+~\tau^\pm~+~{\mathrm{missing~energy}}$',
with two of the 
jets being initiated by $b$-quarks. The largest background to this
signature is most probably due to $q\bar q,gg\to t\bar t gg$ events,
with $t\to bW^+$ and $\bar t \to \bar b \tau^- \bar\nu_\tau$, with
the two gluons yielding low transverse momentum jets in the 
forward and backward directions. 
\item We impose the mentioned transverse momentum and pseudorapidity
constraints on all  six jets: $p_T^j>20$
GeV and $|\eta^j|<5$. Besides, we require that the difference in
pseudorapidity between the two quark/gluon-jets with highest and lowest 
$\eta$-value is larger
than 2: $|\eta_{\mathrm{max}}^j-\eta_{\mathrm{min}}^j| 
\equiv   |\eta^{j_1}-\eta^{j_2}| >  2$.
\item Leptons (electrons and/or muons) are accepted if
$p_T^\ell>20$
GeV and $|\eta^\ell|<2.5$.
\item A pseudorapidity-azimuthal separation between any jet-jet
and jet-lepton pair is imposed: 
$\Delta R\equiv\sqrt{\Delta\eta^2+\Delta\Phi^2}\ge0.7$.
\item A threshold on the  missing transverse energy is enforced too:
$p_T^{\mathrm{miss}}>M_{H^\pm}/2$
(here, we make the assumption that the charged Higgs mass is
already known).
\item We ask that two quark/gluon-jets (among those not satisfying
the last requirement in 1.) reproduce the ${W^\pm}$ mass within
10 GeV: $|M_{{j_3j_4}}-M_{W^\pm}| <  10$ GeV.
\item We further ask that the two above jets reproduce 
the top mass within 25 GeV, if paired with a third
jet: $|M_{{j_3j_4j_5}}-m_{t}| <  25$ GeV.
\item We impose the veto $|M_{{j_6\tau\nu_\tau}}-m_{t}| >  25$ GeV, where
$M_{{j_6\tau\nu_\tau}}$ is the invariant mass obtained by combining the
remaining quark/gluon-jet with the visible $\tau$-momentum and the one
of the parent neutrino, the
latter being reconstructed by adopting the technique outlined in the 
fourth paper of \cite{theory}.  
\item We also cut di-jet invariant masses obtained
from the two jets already identified in 1. which are below the charged
Higgs mass: i.e., $M_{j_1j_2} >  M_{H^\pm}$ (see remark in 5.).
\item Finally, we plot the invariant mass of the four-jet system recoiling 
against the $j_1j_2$ and $\tau^\pm$-neutrino pairs.
\end{enumerate}

Although the number of events of type 1. produced via process (\ref{signal})
can still be sizable at the end of the sequence of cuts in 2.--8., 
and the charged Higgs
mass can be reconstructed rather neatly via
step  9., see Fig.~\ref{fig:mass},
the background from the QCD events  
$q\bar q,gg\to t\bar t gg$ is prohibitive.
Despite having been reduced by several 
orders of magnitude, it overwhelms the 
Higgs resonances completely. It should in fact be noticed that
the integral over the three Higgs curves in Fig.~\ref{fig:mass}, multiplied
by the mentioned annual luminosity (i.e., 100 fb$^{-1}$), yields only
6, 4 and 2 events, in correspondence of $M_{H^\pm}=215, 310$ and
408 GeV, respectively, whereas the background rates sum up to
a total which is typically 1000 times bigger in the vicinity of the peaks.
Besides, the above numbers for the $q q\to q q H^+H^-$
process are obtained for $\tan\beta=40$, value 
for which the product of the branching ratios of the two channels
$H^+\to t\bar b\to b\bar b W^+$ and $H^-\to\tau^-\bar\nu_\tau$
is maximal, within the theoretically preferred $\tan\beta$ interval,
i.e., $\tan\beta\lsim m_t/m_b$.
The situation does not improve substantially for other values of $M_{H^\pm}$
in the heavy mass range
or other, more selective choices of cuts, than those illustrated here.

\section{Summary and conclusions}
\noindent
In summary, we have demonstrated that $H^+H^-$ production can be induced 
at the LHC by three distinct processes: quark-antiquark,  vector-vector and 
gluon-gluon fusion, in order of quantitative importance. 
Whereas the phenomenological relevance of the first and third of these 
modes has been recognised for some time (and stressed again recently), 
that of the second channel
constitutes the novelty of our research. Besides, at the LHC, the $gg$-mode
has been advocated as one of the best ways to probe the 
$hH^+H^-$ and $HH^+H^-$ couplings among Higgs scalars 
and as an effective means to constrain the squark sector of SUSY, as both
neutral Higgs bosons and scalar quarks enter the virtual stages
of the production process. Hence, in suppressing the background 
to $gg\to H^+H^-$, special care
has to be adopted in dealing not only with the $q\bar q\to
H^+H^-$ mode, but also with the $q q\to q qH^+H^-$ channel.
Finally, the EW vector-vector fusion reaction has  been tested as
a possible detection mode of heavy charged Higgs bosons, as a 
complement to the leading $gg\to \bar t b H^+$ channel, the former
covering the 
$3\lsim\tan\beta\lsim10$ window, where the detection potential
of the latter is seriously hampered by a steeply falling production
rate (with a minimum at $\tan\beta=7$). Despite the independence
of $q q\to q qH^+H^-$ from $\tan\beta$ could allow for a
prompt estimate of $M_{H^\pm}$ and the consequent
mass resonance selection can be made viable,
the background from QCD induced events of the type $q\bar q,gg\to 
t\bar t gg$  is prohibitively large, at least in the 
channel $H^+H^-\to `4~{\mathrm{jets}}~+~{\tau^\pm}~+$
missing energy', including the detection of the two quark-jets
produced in association with the Higgs boson pair. Consequently,
we expect the measurement of the triple Higgs couplings entering 
process (\ref{signal}) via graphs 15,16 and 18 in Fig.~\ref{fig:graphs}
to be extremely difficult. As forward-jet tagging
has proved to be a crucial ingredient of our analysis 
in reducing the QCD noise, and if one also recalls Fig.~\ref{fig:ggHH},
showing the dominance of $qq\to qqH^+H^-$ over $gg\to H^+H^-$, similar
conclusions should apply to the case of the gluon-induced reaction. Finally,
even in case signals of the $q\bar q \to H^+H^-$ process can be 
detected (again, see Fig.~\ref{fig:ggHH}), it should be remembered that
sizable effects of Higgs self-couplings enter here only via s-channel
$b\bar b$-annihilation, which is a small component of the 
total $q\bar q$-induced 
rate, further considering that the $b\bar b\to H^+H^-$ cross section is
certainly over-estimated by the current sets of $b$-quark PDFs, given
that the use of the more realistic $gg\to b\bar b H^+H^-$ 
partonic scattering yields a rate which is about an order of magnitude
smaller, whenever $M_{H^\pm}\gsim m_t$ (recall Fig.~\ref{fig:bbHH}).

Indeed, if the processes and signature 
that we have chosen are to be useful in searching for 
pairs of charged Higgs bosons at the LHC, far better cuts than those designed
here will need to be devised. An alternative, cleaner detection mode
could be $H^+H^-\to \tau^+\nu_\tau\tau^-\bar\nu_\tau$, however,
this requires more realistic simulations (including double $\tau^\pm$
reconstruction in a real detector environment) than those that can 
be carried out in a parton level study. Ultimately, processes of the
type $q\bar q\to H^+H^-$, $gg\to H^+H^-$ and $qq\to qq H^+H^-$ are
primary candidates to benefit from a possible tenfold LHC luminosity upgrade, 
the so-called SLHC \cite{SLHC},
given their rather small production rates at the standard LHC in general.

\section*{Acknowledgements}
We are grateful to PPARC for financial support and to the Theory
Group at the Rutherford Appleton Laboratory (RAL) for hospitality
during the early stages of this work. We also thank Kosuke Odagiri
for suggesting the topic of this research, for
innumerable helpful discussions and for numerical comparisons as well.

\vfill\clearpage\thispagestyle{empty}

\begin{figure}[!t]
\begin{center}
~\hskip-6.5cm{\epsfig{file=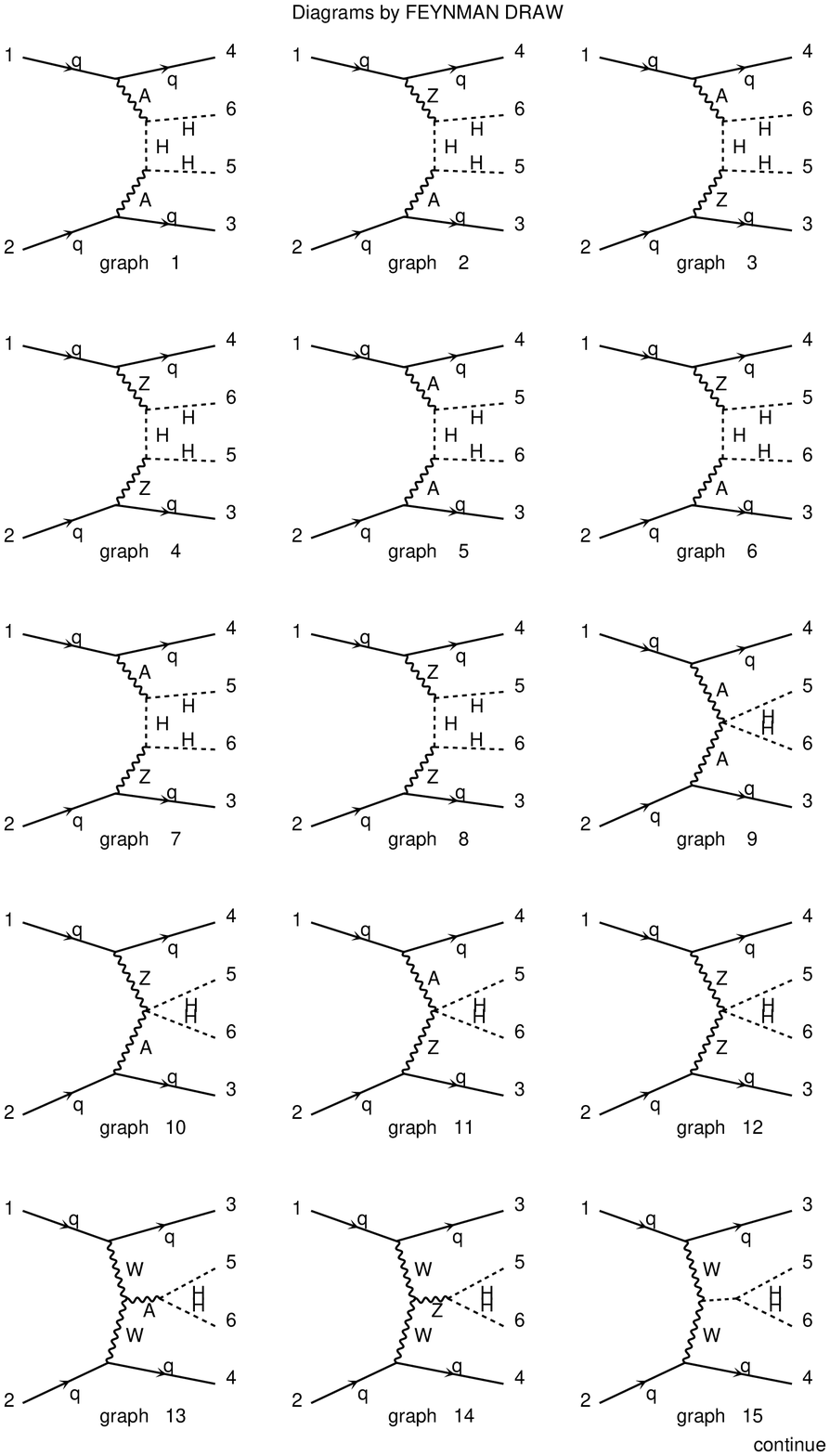,height=23cm,angle=0}}
\end{center}
\end{figure}

\vfill\clearpage\thispagestyle{empty}

\begin{figure}[!t]
\begin{center}
~\hskip-6.5cm{\epsfig{file=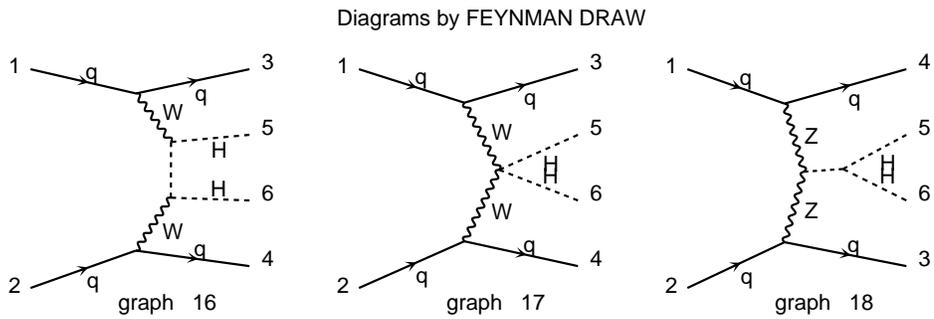,height=23cm,angle=0}}
\end{center}
\begin{center}
\vskip-18truecm
\caption{Feynman diagrams at tree level for process
 (\ref{signal}). The labels {\tt q, A, Z, W and H} refer
to a (anti)quark, $\gamma$, $Z$, $W^\pm$ and $H^\pm$ boson,
respectively, whereas an unlabelled, internal dashed line represents
a summation over $H, h$ and $A$ boson propagators.}
\label{fig:graphs}
\end{center}
\end{figure}

\vfill\clearpage\thispagestyle{empty}

\begin{figure}[!t]
\begin{center}
~{\epsfig{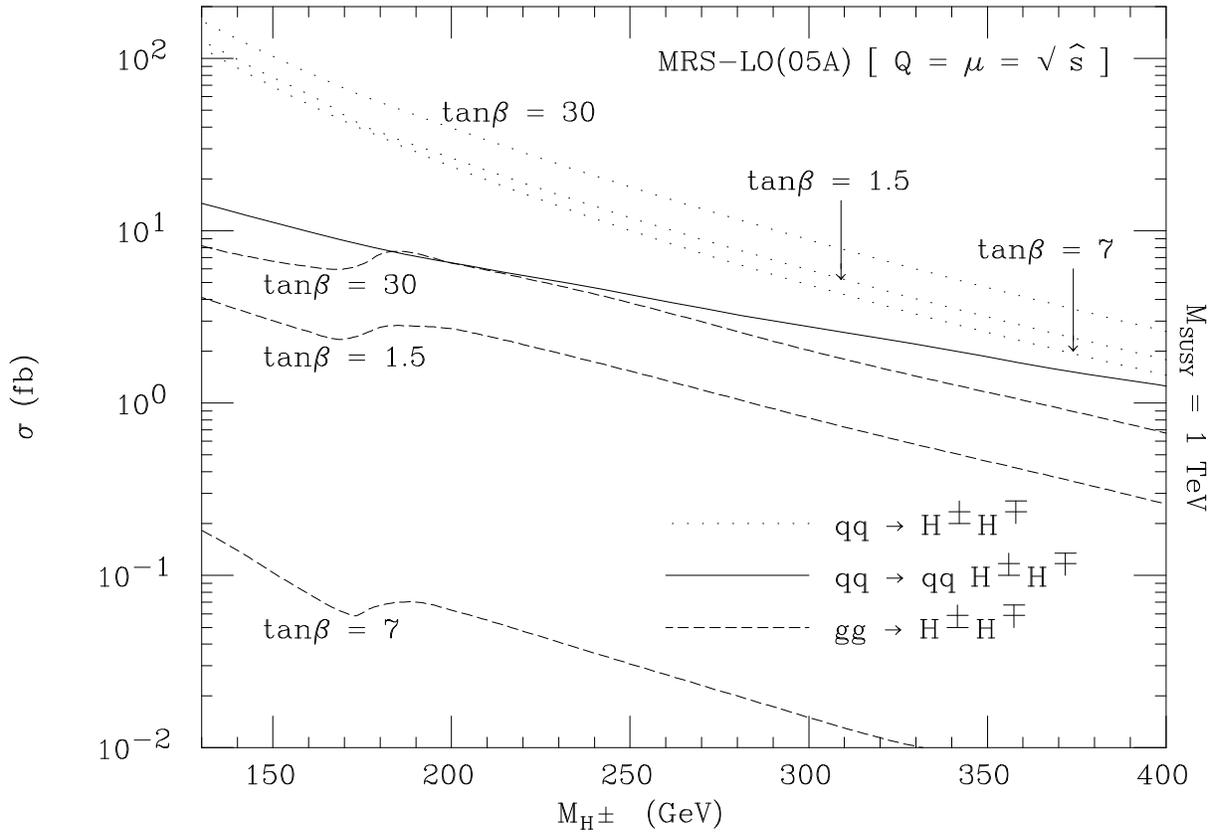}}
\end{center}
\begin{center}
\caption{Cross section in femtobarns at the LHC for the following 
$H^+H^-$ production processes discussed in the text:
$q\bar q\to H^+H^-$ (including the $b\bar b$ contribution), 
$gg\to H^+H^-$ and $q q\to q qH^+H^-$, 
 for $\tan\beta=1.5,7$ and 30.
In the last process, there is no visible $\tan\beta$ dependence.}
\label{fig:ggHH}
\end{center}
\end{figure}

\vfill\clearpage\thispagestyle{empty}

\begin{figure}[!t]
\begin{center}
~\hskip-6.5cm{\epsfig{file=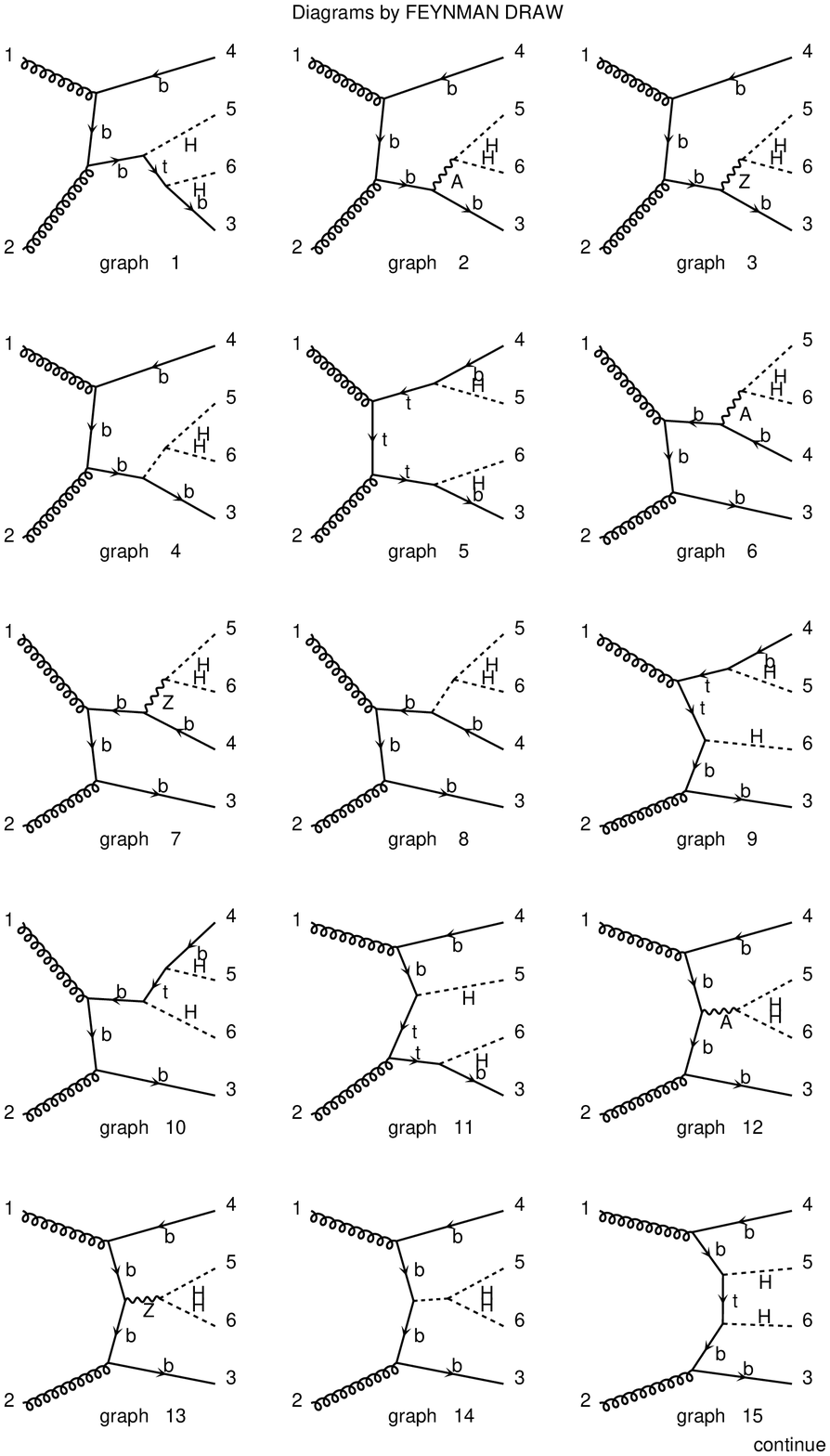,height=23cm,angle=0}}
\end{center}
\end{figure}

\vfill\clearpage\thispagestyle{empty}

\begin{figure}[!t]
\begin{center}
~\hskip-6.5cm{\epsfig{file=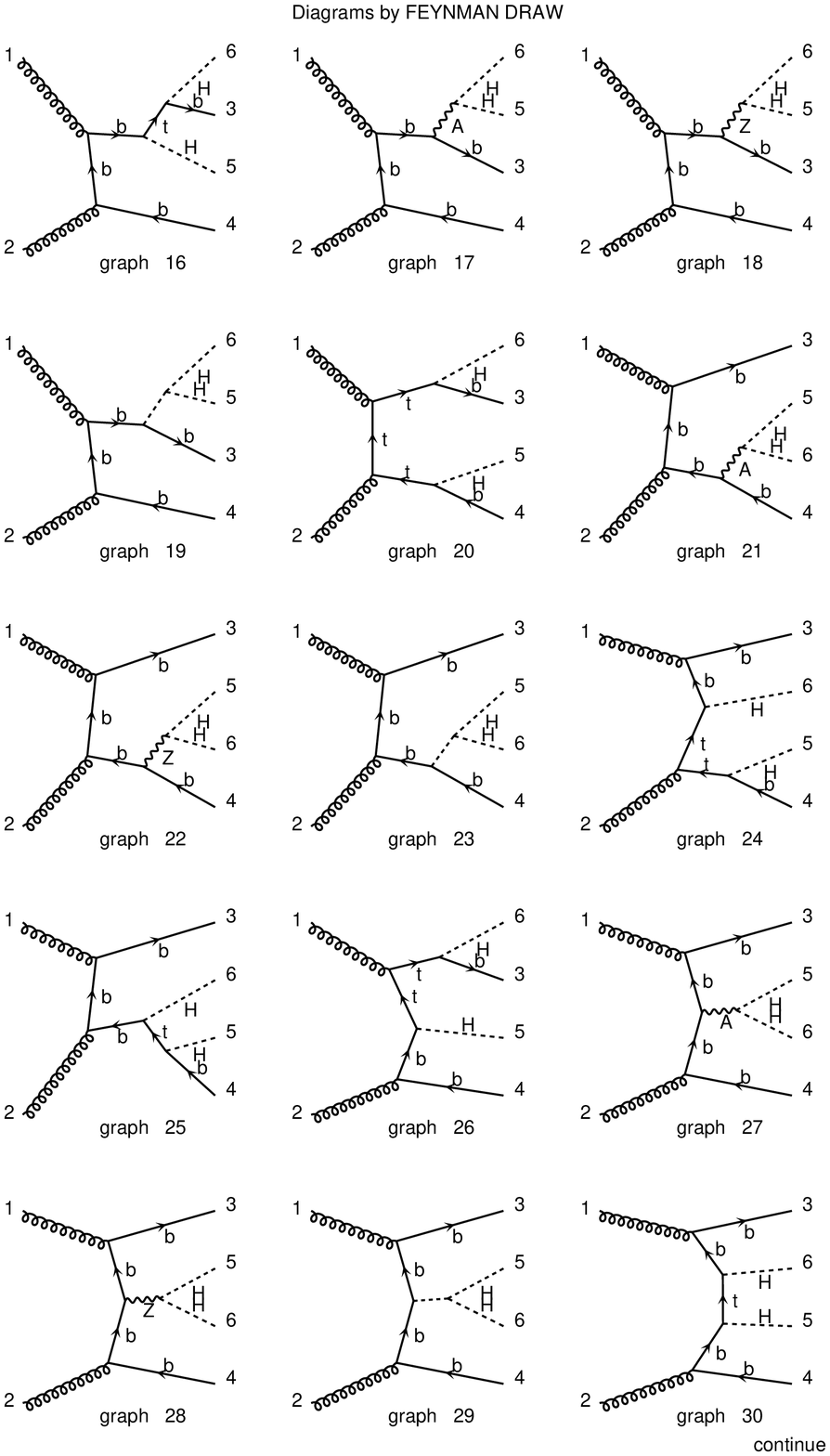,height=23cm,angle=0}}
\end{center}
\end{figure}

\vfill\clearpage\thispagestyle{empty}

\begin{figure}[!t]
\begin{center}
~\hskip-6.5cm{\epsfig{file=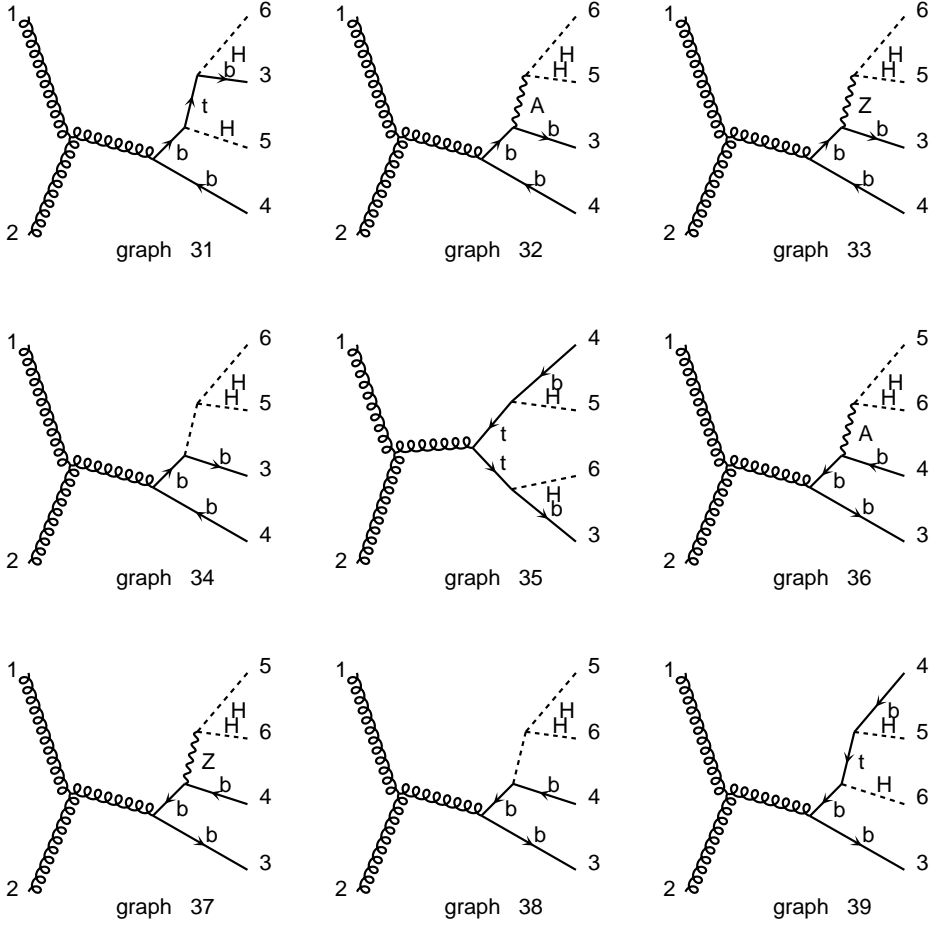,height=23cm,angle=0}}
\end{center}
\begin{center}
\vskip-10truecm
\caption{Feynman diagrams at tree level for process
 (\ref{ggbbHH}). The labels {\tt b, t, A, Z, W and H} refer
to a $b,t$-(anti)quark, $\gamma$, $Z$, $W^\pm$ and $H^\pm$ boson,
respectively, whereas an unlabelled, (internal)[external] 
(dashed)[helical] line represents
a (summation over $H$ and $h$ boson propagators)[gluon].}
\label{fig:ggbbHH}
\end{center}
\end{figure}

\vfill\clearpage\thispagestyle{empty}

\begin{figure}[!t]
\begin{center}
~{\epsfig{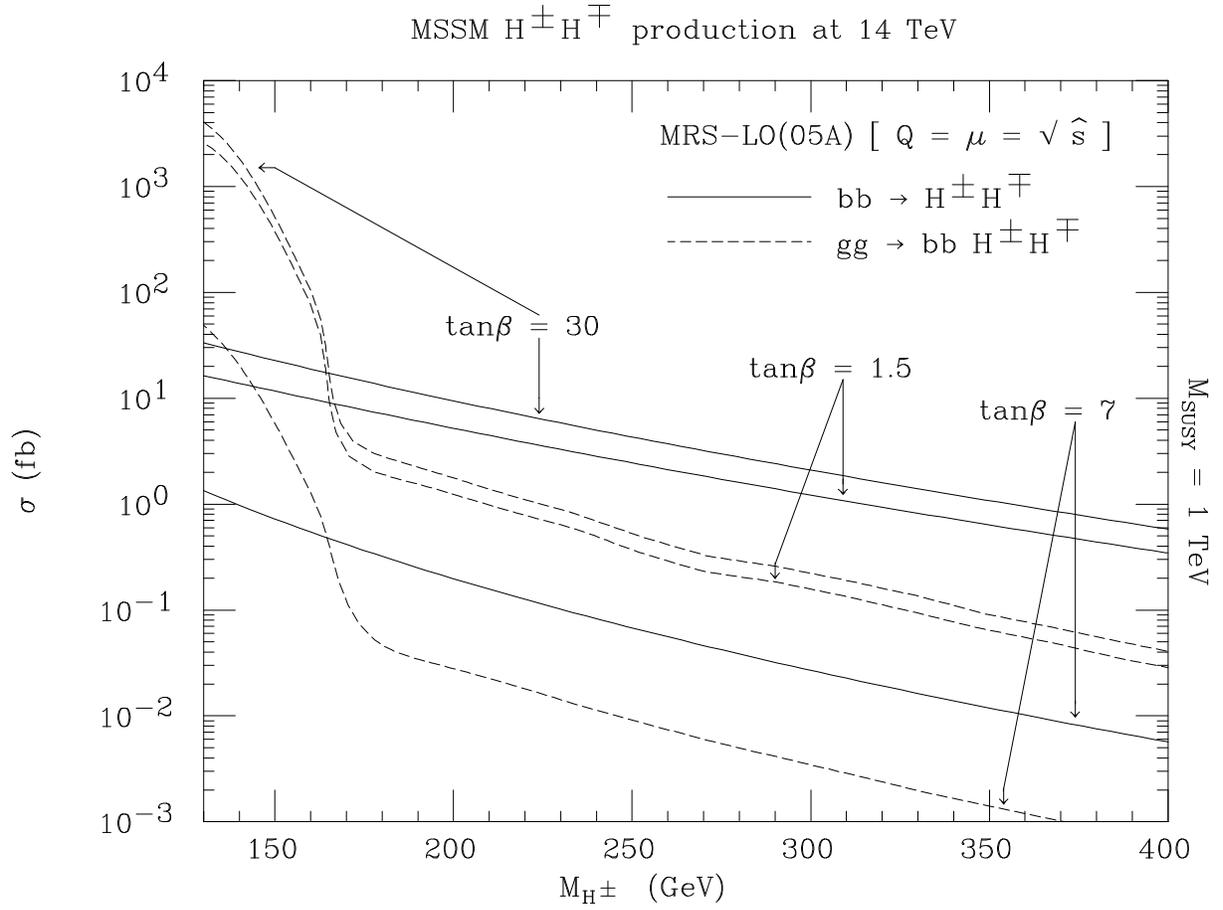}}
\end{center}
\begin{center}
\caption{Cross section in femtobarns at the LHC for the following 
$H^+H^-$ production processes discussed in the text:
$b\bar b\to H^+H^-$ and $g g\to b\bar bH^+H^-$, 
for $\tan\beta=1.5,7$ and 30.}
\label{fig:bbHH}
\end{center}
\end{figure}

\vfill\clearpage\thispagestyle{empty}

\begin{figure}[!t]
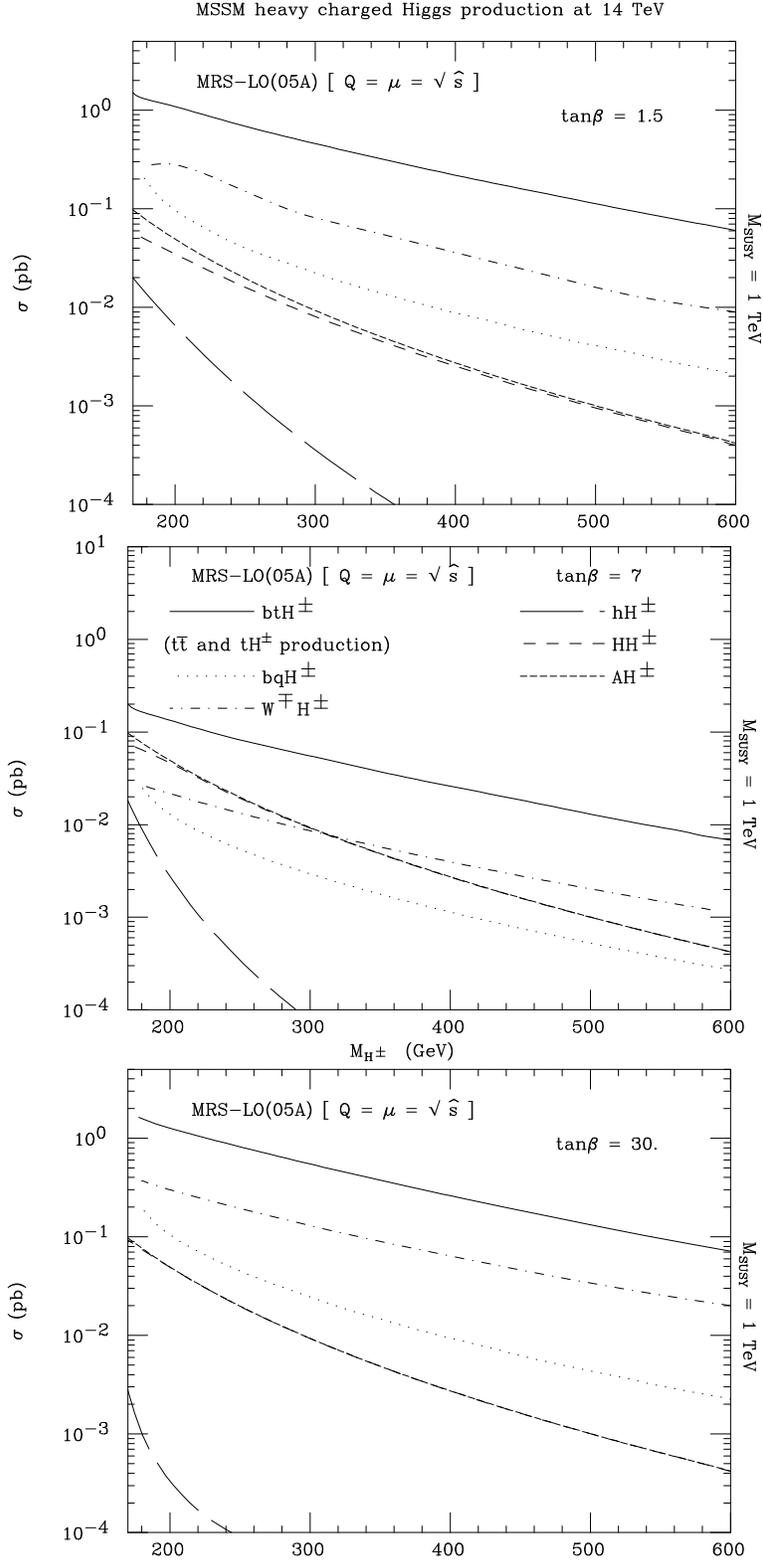

\begin{center}
~{\epsfig{file=charged_a.ps,height=10cm,angle=90}}
~{\epsfig{file=charged_c.ps,height=10cm,angle=90}}
~{\epsfig{file=charged_b.ps,height=10cm,angle=90}}
\end{center}
\begin{center}
\vskip-0.5cm
\caption{Cross sections in picobarns at the LHC for the
production mechanisms of a single charged Higgs boson, 
 for $\tan\beta=1.5$ (top), 7 (middle) and 30 (bottom).
(The $q\bar q'\to \Phi H^\pm$ rates, with $\Phi=h,A$, visually
coincide for $\tan\beta=30$.)}
\label{fig:comparison}
\end{center}
\end{figure}

\vfill\clearpage\thispagestyle{empty}

\begin{figure}[!t]
\begin{center}
~{\epsfig{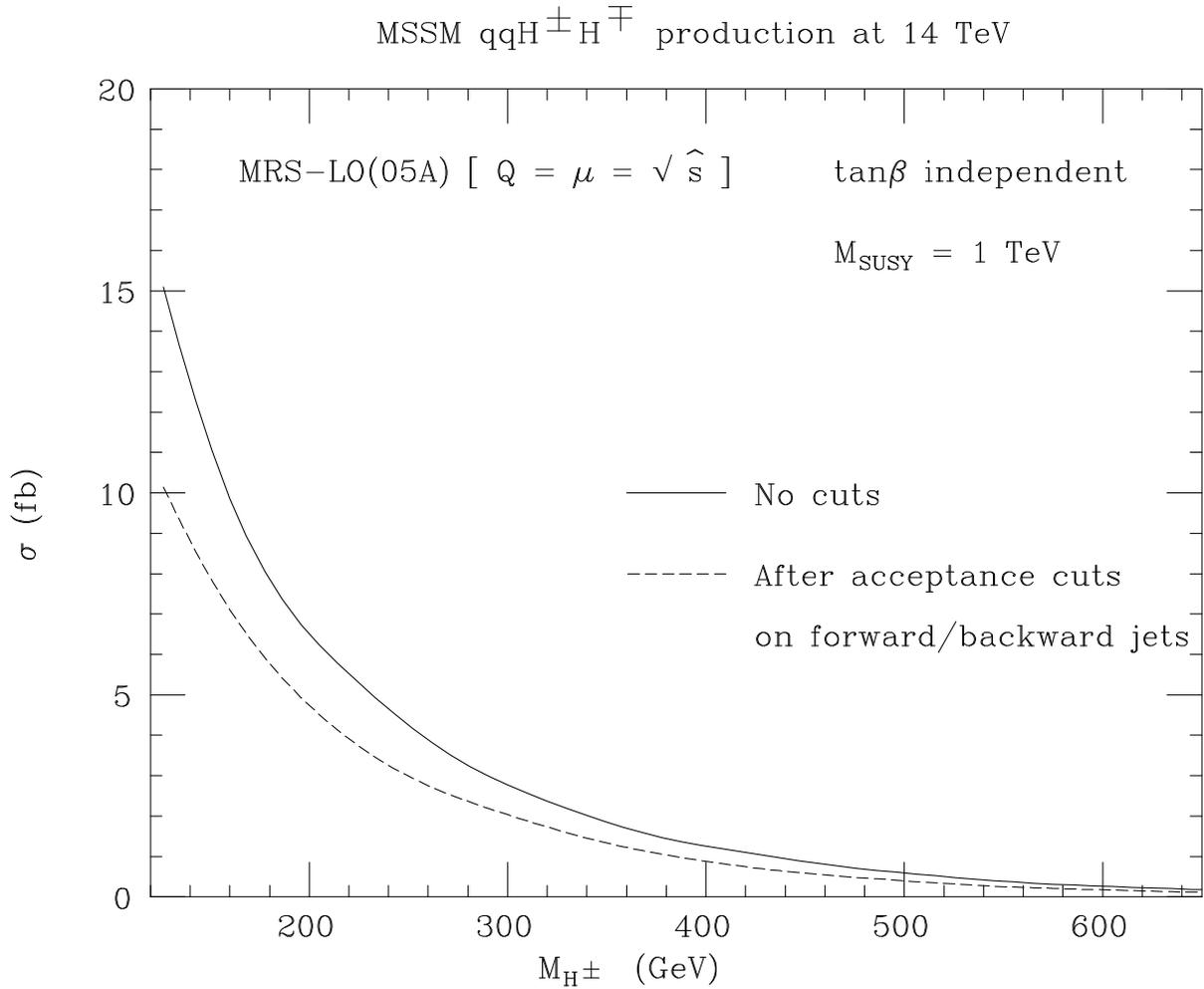}}
\end{center}
\begin{center}
\caption{Cross sections in femtobarns at the LHC for process
 (\ref{signal}), without and with the following acceptance cuts on the 
forward jets: $p_T^j>20$ GeV and $|\eta^j|<5$. Here, there is no
visible $\tan\beta$ dependence.}
\label{fig:all}
\end{center}
\end{figure}

\vfill\clearpage\thispagestyle{empty}

\begin{figure}[!t]
\begin{center}
~{\epsfig{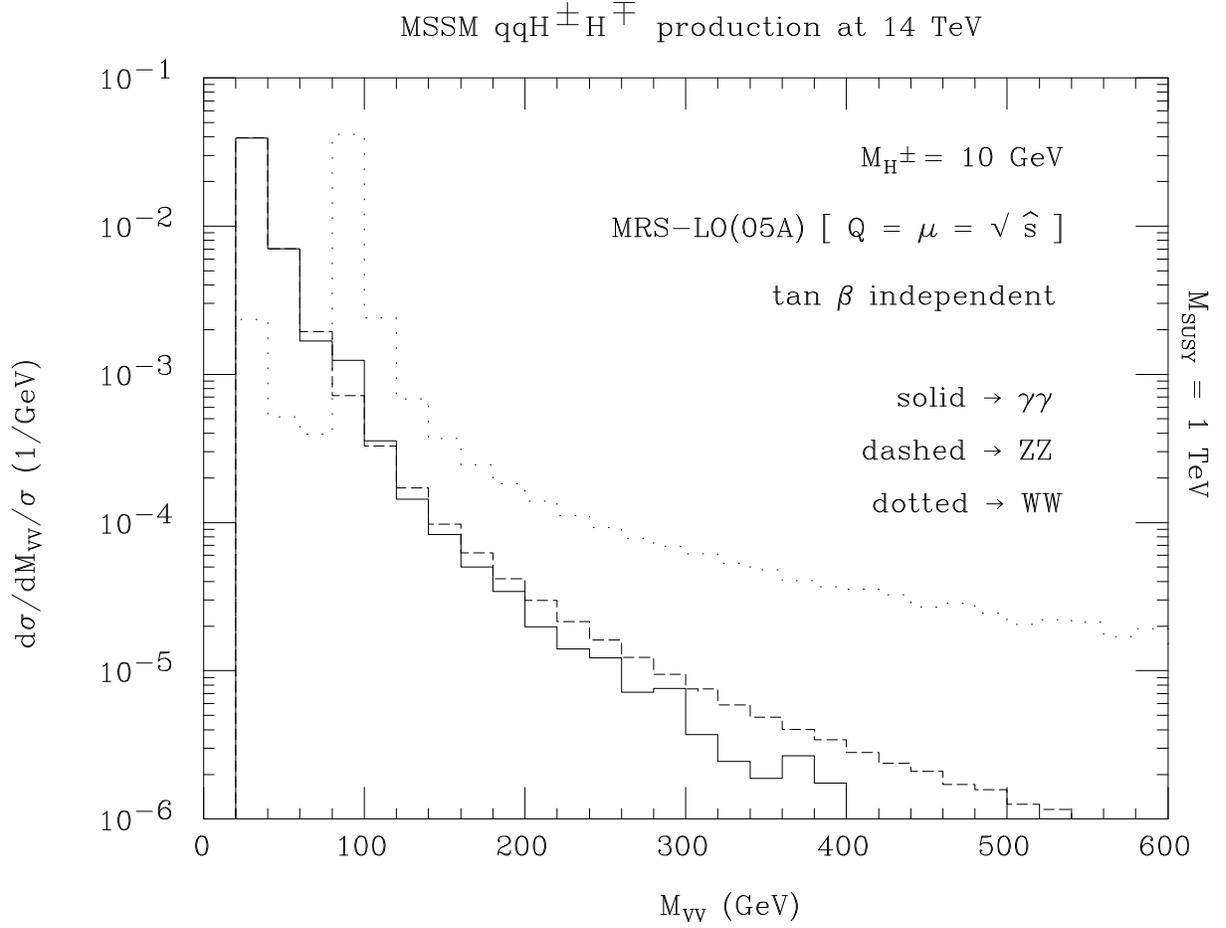}}
\end{center}
\begin{center}
\caption{Mass distribution of the vector-vector system recoiling
against the forward/backward jets, before any cuts, in case
of the the $\gamma\gamma,ZZ$ and $W^+W^-$ mediated graphs, out of the
full set of process (\ref{signal}), for the `illustrative'
value $M_{H^\pm}$=10 GeV.
Here, there is no visible $\tan\beta$ dependence.}
\label{fig:contributions}
\end{center}
\end{figure}

\vfill\clearpage\thispagestyle{empty}

\begin{figure}[!t]
\begin{center}
~{\epsfig{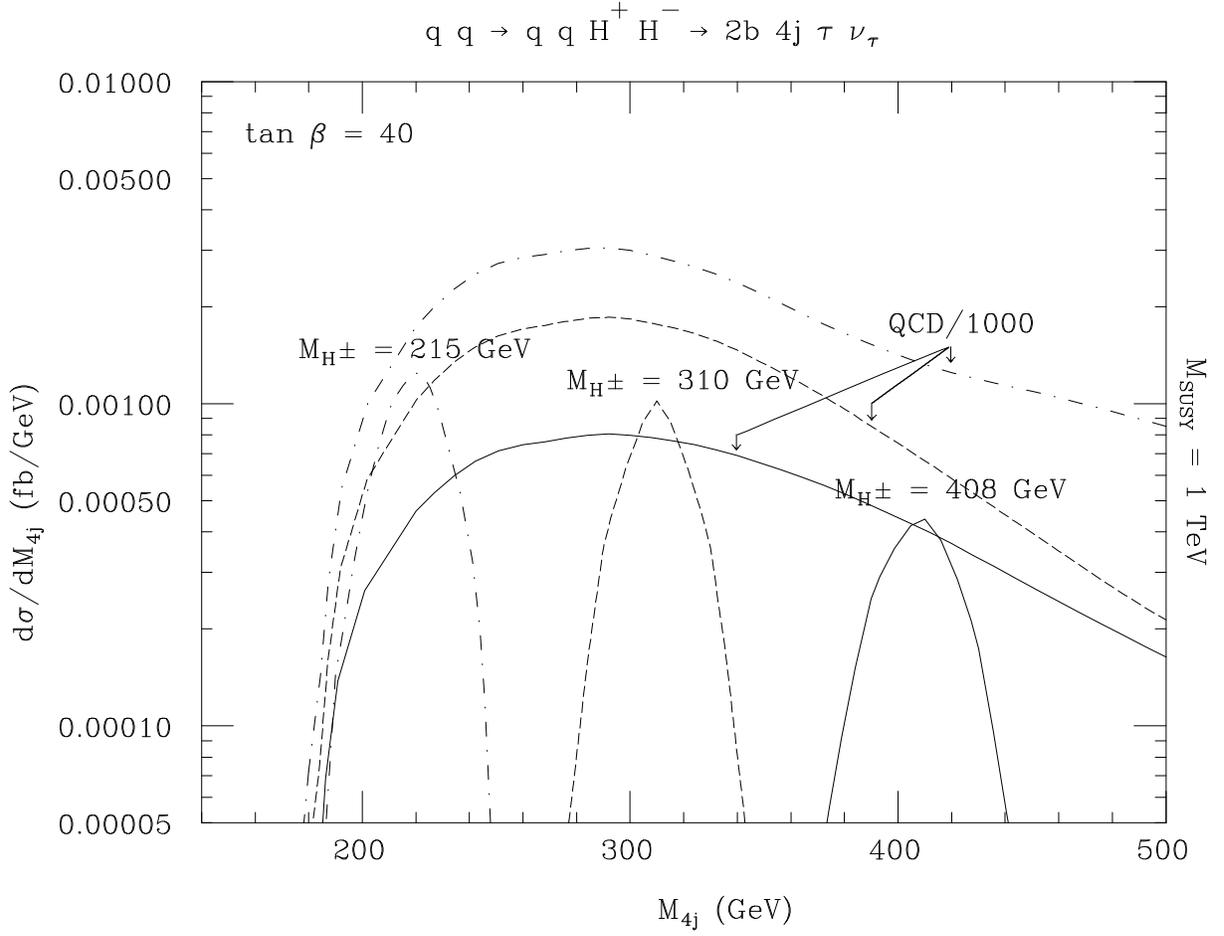}}
\end{center}
\begin{center}
\caption{Mass distribution of the four-jet system recoiling
against the $j_1j_2$ and $\tau^\pm$-neutrino pairs,
after the selection described in points 2.--8.,
in case of process (\ref{signal}) and of the QCD background
$q\bar q,gg\to t\bar t gg$, for three choices of $M_{H^\pm}$ in the heavy 
mass range and $\tan\beta=40$,  in the channel
`$6~{\mathrm{jets}}~+~\tau^\pm~+~{\mathrm{missing~energy}}$'.}
\label{fig:mass}
\end{center}
\end{figure}

\end{document}